\documentclass{vgtc}                          

\graphicspath{{figures/}{pictures/}{images/}{./}} 
\usepackage{amsmath}

\usepackage{times}                     

\usepackage{tabu}                      
\usepackage{booktabs}                  
\usepackage{lipsum}                    
\usepackage{mwe}                       

\usepackage{mathptmx}  
\usepackage{enumitem}

\usepackage{cite}

\vgtccategory{Research}

\vgtcinsertpkg

\usepackage{hyperref}

\usepackage{microtype}
\usepackage{etoolbox}
\usepackage{booktabs}
\usepackage{refcount}
\usepackage{totpages}
\usepackage{environ}
\usepackage{setspace}
\usepackage{textcase}
\usepackage{verbatim}
\usepackage{xspace}
\usepackage{hyperref} 

\usepackage{url}
\newcommand \systemname{\emph{AdaptiveCoPilot}\xspace}
\def \projectName {{AdaptiveCoPilot}}

\newcommand{\revision}[1]{{\color{black}{#1}}}

\title{\projectName{}: Design and Testing of a NeuroAdaptive LLM Cockpit Guidance System in both Novice and Expert Pilots} 

\author{Shaoyue Wen\thanks{e-mail: sw6352@nyu.edu}\\ %
        \scriptsize New York University %
\and Michael Middleton*\thanks{e-mail: Michael.Middleton@ngc.com}\\ %
     \parbox{1.4in}{\scriptsize \centering New York University \& Northrop Grumman} %
\and Songming Ping\\ %
     \parbox{1.4in}{\scriptsize \centering  Imperial college}
\and Nayan N Chawla\\ %
     \scriptsize \centering Virginia Tech
\and Guande Wu\\ %
     \scriptsize \centering New York University
\and Bradley S Feest\\ %
     \scriptsize \centering Northrop Grumman
\and Chihab Nadri\\ %
     \scriptsize \centering University of Illinois, Urbana-Champaign
\and Yunmei Liu\\ %
     \scriptsize \centering University of Louisville
\and David Kaber\\ %
     \scriptsize \centering University of Florida
\and Maryam Zahabi\\ %
     \scriptsize \centering Texas AM University
\and Ryan P. McMahan\\ %
     \scriptsize \centering Virginia Tech
\and Sonia Castelo\\ %
     \scriptsize \centering New York University
\and Ryan Mckendrick**\thanks{e-mail: ryan.mckendrick@ngc.com}\\ %
     \scriptsize \centering Northrop Grumman
\and Jing Qian**\thanks{e-mail: jq2267@nyu.edu}\\ %
     \scriptsize \centering New York University
\and Cl\'{a}udio T. Silva**\thanks{e-mail: csilva@nyu.edu}\\ %
     \scriptsize \centering New York University}

\newcommand\myparagraph[1]{\vspace{4pt} \noindent \textbf{#1.}}

\newcommand{\participantquote}[1]{``\textit{#1}''}

\teaser{
  \centering
  \includegraphics[width=\linewidth]{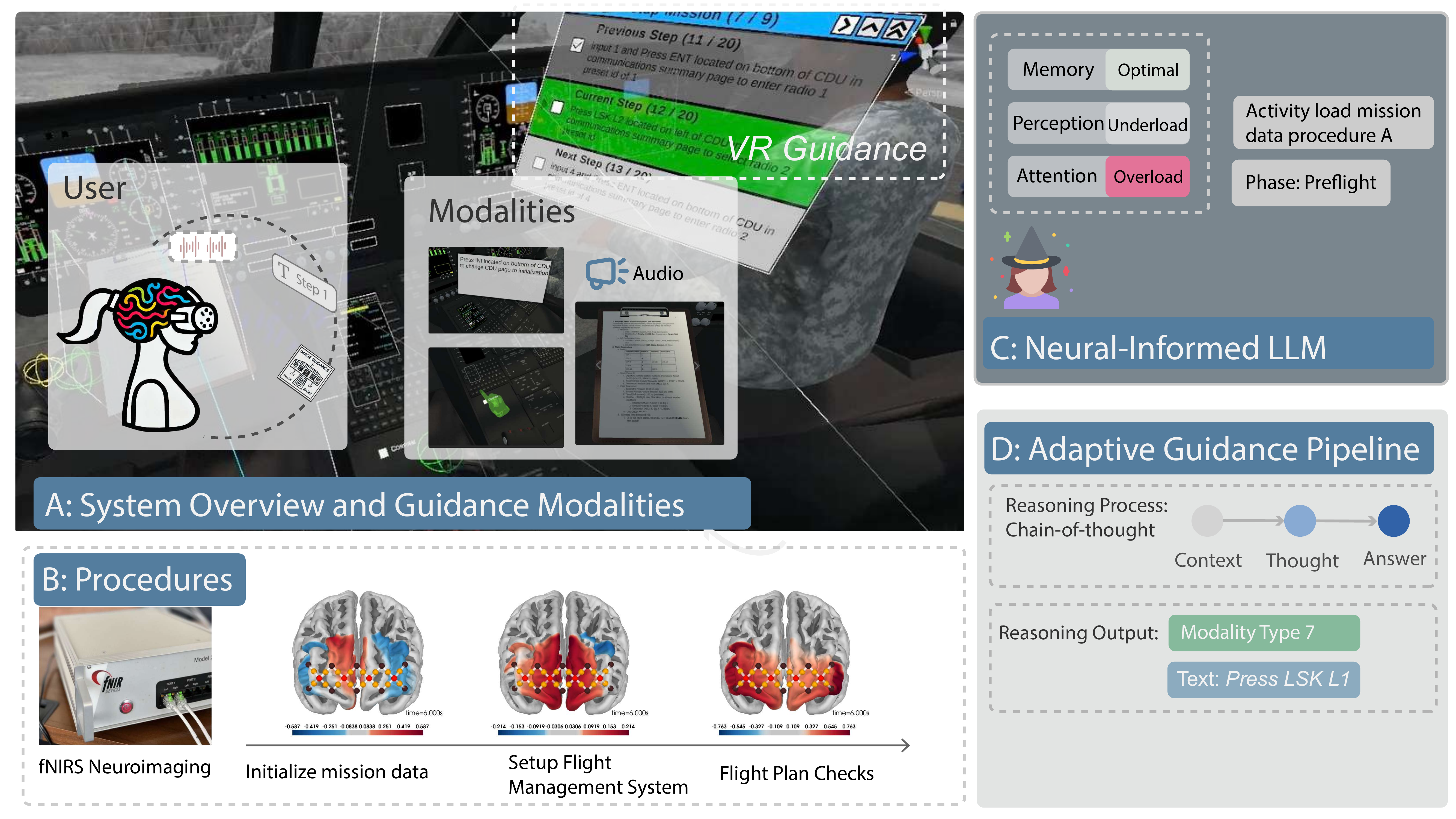}
  \caption{  \textbf{A}. \systemname  is a neuroadaptive guidance system for pilot virtual reality preflight checklist training. \textbf{B}. Cognitive workload state is assessed through fNIRS neuroimaging classification models. \textbf{C}. \systemname combines cognitive workload states, behavioral information, and carefully constructed adaptive rules into a context aware LLM. \textbf{D}. The Neural-Informed LLM reasons over the input to inform modality and information load adaptations with the goal of maintaining a pilots optimal workload.}
  \label{fig:teaser}
}

\abstract{
Pilots operating modern cockpits often face high cognitive demands due to complex interfaces and multitasking requirements, which can lead to overload and decreased performance. This study introduces \systemname, a neuroadaptive guidance system that adapts visual, auditory, and textual cues in real time based on the pilot's cognitive workload, measured via functional Near-Infrared Spectroscopy (fNIRS). A formative study with expert pilots (N=3) identified adaptive rules for modality switching and information load adjustments during preflight tasks. These insights informed the design of \systemname, which integrates cognitive state assessments, behavioral data, and adaptive strategies within a context-aware Large Language Model (LLM). The system was evaluated in a virtual reality (VR) simulated cockpit with licensed pilots (N=8), comparing its performance against baseline and random feedback conditions. The results indicate that the pilots using \systemname exhibited higher rates of optimal cognitive load states on the facets of working memory and perception, along with reduced task completion times. Based on the formative study, experimental findings, qualitative interviews, we propose a set of strategies for future development of neuroadaptive pilot guidance systems and highlight the potential of neuroadaptive systems to enhance pilot performance and safety in aviation environments.

}

\keywords{ Aviation in virtual reality, Adaptive user interface}

\begin{document}

\maketitle

\section{Introduction}

\revision{Modern cockpits have grown in both complexity and features, as such they require increasingly strong operational skill and knowledge \cite{socha_pilots_2020}. Pilots often engage in multi-tasking which may saturate their cognitive abilities. Prolonged cognitive load saturation can also induce fatigue along with decreased performance, leading to errors and potential aircraft incidents \cite{barter_aspects_1993, cao2024computational}. Advancements in aviation training and guidance systems have not kept pace with the complexity of modern cockpits. Paper checklists are the primary form of aviation task guidance \cite{degani_human_1991} and Crew Resource Management (CRM) training, although effective at reducing errors, has changed little since 1981 \cite{helmreich_evolution_1999} .}



\revision{CRM training for aviation is defined as training to use hardware, software, and humans effectively to minimize aviation errors \cite{jensen_pilot_2017}. Mnemonic strategies such as ``aviate–navigate–communicate'' are taught to pilots in CRM courses as a hierarchical memorization technique to assign priorities and regulate workload in high-stress emergencies \cite{morris_pilot_2006}. However, these strategies themselves require attentional resources. Further, the many step procedures found in aviation checklists~\cite{barron_multitasking_2017}, are difficult to memorize through these techniques.} \revision{Recently, brain-computer interfaces have shown potential in measuring and addressing areas of high stress and cognivive demand in aviation environments \cite{zhang_effect_2024, dehais_mental_2023}.
} 

This research explores the benefits of \systemname{}, an \textbf{adaptive, multimodal feedback} system implemented in a virtual reality (VR) cockpit that provides real-time guidance based on pilots' cognitive states and performance. The system uses functional Near-Infrared Spectroscopy (fNIRS) to measure cognitive states across three facets—working memory, perception, and attention—and adaptively adjusts the modality and content of visual and auditory feedback. 
A formative study with three pilots \revision{found the specific areas of} high cognitive demands of aircraft operation and the challenges in selecting appropriate modality and content for feedback. 
The results from this study informed the development of neuroadaptive strategies aimed at maintaining optimal workload states across different cognitive facets. An eight-pilot case study provided initial evidence of the system's effectiveness in managing cognitive load. Additionally, after completion of the study interviews with pilots from diverse backgrounds—including recreational, fighter, and expert Black Hawk pilots—offered valuable perspectives on the role of automated guidance and the benefits of the neuroadaptive system.

This research makes the following contributions:
\begin{itemize}

\vspace{-0.08in}

    \item \systemname{}, a system  adapts the content and guidance modality via users' cognitive states, providing visual, auditory, and text-based guidance with \textit{adaptive rules} summarized from existing theories. 

\vspace{-0.08in}

    
    \item A case study on real pilots testing the efficacy of \systemname{} through a VR-simulated preflight task, comparing baseline, random and adaptive conditions. Thus providing a functional proof-of-concept for generalized real-time neuroadpative guidance during pre-flight.

\vspace{-0.08in}
    
    \item The insights gained from the qualitative evaluation are synthesized into strategies for designing and testing future AI-adaptive assistants. 
\end{itemize}

\section{Related Work}
\label{sec:related}
We review the use of VR in high-fidelity simulations, methods to measure cognitive workload, and the development of neuroadaptive systems for adaptive feedback based on real-time physiological data.

\myparagraph{VR for Testing New Technology with High Ecological Validity}
VR allows for the recreation of complex environmental contexts, offering immersive experiences that enhance ecological validity in testing and training scenarios~\cite{hu_effects_2020, hu_evaluating_2020}. In aviation, VR simulators provide pilots with realistic training environments where they can practice procedures and respond to simulated emergencies without the risks associated with real flights~\cite{aguilar2023design}. Similarly, VR has been used in military simulations to train personnel in realistic combat scenarios, improving readiness and decision-making skills~\cite{hunter2021, muhlberger2001}. This makes VR ideal for testing prototype experimental systems due to relative safety and its effectiveness.

\vspace{-0.05in}
\myparagraph{Measuring Cognitive Workload}
Cognitive workload is the mental effort required to perform a task and is influenced by the interaction between task demands and an individual's cognitive capabilities~\cite{young_state_2015, gartner1979, moray2013, thorpe_systematic_2020}. Our system focuses on cognitive capabilities in working memory, perception, and attention, as each of these cognitive facets has inherent processing limitations \cite{wickens_multiple_2008}.

Since cognitive workload is not directly observable, it is commonly estimated using three primary methods. \emph{(1) Task Performance Measures:} Evaluating performance on a primary or concurrent secondary task to infer cognitive load~\cite{moray2013}. \emph{(2) Subjective Evaluations:} Using self-reported assessments such as the NASA Task Load Index (NASA-TLX)~\cite{hart1988} and the Modified Cooper-Harper Scale~\cite{hill1992}. These methods can suffer from self-reporting biases and can often be misleading if not analyzed, accounting for individual differences in participant bias~\cite{mckendrick_deeper_2018}. \emph{(3) Physiological Measures:} Tracking biological responses such as respiration, eye movements, and brain signals which represent the underlying causal components of cognitive processing but are often noisy and difficult to interpret~\cite{charles2019}.

\vspace{-0.05in}
\myparagraph{Real-Time Physiological Measurement of Workload}
Physiological sensors and algorithmic classifiers can be used to provide unobtrusive estimates of cognitive workload in real-time. Monitoring cortical hemodynamics via the prefrontal cortex—critical for processing and control of working memory, perception, and attention~\cite{funahashi_working_2017, helfrich_prefrontal_2017} has been demonstrated as a viable method for estimating and predicting the effects of task demands and their interaction with individual difference on these cognitive facets of workload~\cite{mckendrick_theories_2019, mckendrick_cognitive_2019, pinti_present_2020}. A functional Near-Infrared Spectroscopy (fNIRS) device is a minimally invasive method for monitoring cortical hemodynamics and can be quickly set up under a VR headset. This method is also robust to movement~\cite{ayaz_using_2011} and commonly used in pilot workload studies~\cite{castelo2024HuBar, DBLP:conf/mc/AuerGRJ21, DBLP:journals/ctw/OberhauserD17}. \revision{Similarily, pupilometry-based approches are also used to determine the cognitive load via changes in pupil size and are robust to task complexity~\cite{duchowski_lowhigh_2020}. However, we opted for fNIRS due to the need to measure cortical activity relevant to both overload and optimal load conditions directly from the central nervous system. Moreover, to our knowledge, there is currently no open-source real-time pupilometry pipeline that robustly differentiates distinct processing networks in situ, making fNIRS more suitable for this study.}

\vspace{-0.05in}
\myparagraph{Importance of Isolating Cognitive Facets and Considering Non-linear Dynamics}
When measuring workload, isolating the specific cognitive facet driving state changes \revision{(working memory, attention, perception)} is crucial, as it influences how errors should be addressed. State transitions (e.g., from underload to optimal load) are more significant than minor fluctuations corresponding to slight task demand changes \cite{mckendrick_deeper_2018}. This is because by identifying the constructs of underload and overload in neurological data we can make explicit assumptions about performance. 

Understanding the relationship between cognitive load states and performance is essential for designing effective adaptive systems. \revision{McKendrick et al. investigated how varying levels of working memory load affect prefrontal hemodynamics, revealing a cubic relationship rather than a simple linear slope \cite{mckendrick_cognitive_2019}. Specifically, as load increases from underload toward an optimal point, neural activity tends to increase in a predictable manner, but when load surpasses an individual’s capacity (overload), activity attenuates—and performance suffers. Such non-linear dynamics resonate with existing theoretical frameworks. The Yerkes-Dodson law posits a curvilinear relationship between arousal (or workload) and performance \cite{corbett_law_2015}, while Wickens’ Multiple Resource Theory highlights that cognitive resources are limited but allocated flexibly across tasks \cite{wickens_multiple_2008}.} 


\vspace{-0.05in}
\myparagraph{Using Large Language Models for Adaptive Guidance}
LLMs have been increasingly used for high-stakes applications such as bio-medicine~\cite{jin2023genegpt, chen2023evaluation, biswas2023errors} and the judiciary~\cite{jovanovic2020chatbots}. 
\revision{Recently, LLMs have also demonstrated the ability to adapt to new tasks in the HCI domain by leveraging few-shot learning ~\cite{DBLP:conf/chi/WuQQCRS24, DBLP:conf/chi/ZhengWSMLM23}. Following this approach, we prompt Microsoft’s PHI-3 LLM with domain knowledge, adaptive rules and relevant psychological theory, aiming to enable context-aware guidance actions~\cite{auer2021vr}.}

\vspace{-0.05in}
\myparagraph{Concept of Neuroadaptivity}
Neuroadaptive systems adapt to a user's cognitive state, attempting to enhance task performance and manage cognitive strain \cite{zander_neuroadaptive_2016, andreessen_toward_2021}. While augmented visualizations can aid task performance, they are often rigidly applied and do not account for fluctuations in cognitive load. Implementing neuroadaptive strategies involves adjusting system behavior based on \emph{real-time} physiological measurements to maintain optimal workload states.

\revision{However, such dynamic adaptivity can have downsides. One concern is unpredictability: users may find it disruptive or confusing when system behavior changes unexpectedly. Additionally, frequent or overly sensitive adaptations could undermine user autonomy. Finally, imperfect system performance could hinder user acceptance of the system.
As such, adaptive content must be carefully selected and controlled to mitigate these concerns. The following \textbf{(B1-B4)} are pinpointed areas of adaptivity and cognitive load literature that are important to address in an effective neuoradaptive system.
}

\vspace{-0.05in}
\myparagraph{B1. Cognitive Load Management of Information Bandwidth}
As previously noted, by accessing cognitive load in real-time and adjusting the amount of information presented cognitive load can be altered~\cite{haapalainen2010cognitive, lindlbauer2019context}. During high-load tasks, adaptive systems can limit non-critical information, allowing pilots to focus on essential data without being overwhelmed \cite{healey2005detecting}. Such adaptive feedback systems may prove crucial for maintaining optimal performance and preventing cognitive fatigue.

\vspace{-0.05in}
\myparagraph{B2. Task-Specific Information Presentation}
Additionally, adaptive rules can adjust information delivery to be task-specific, ensuring pilots receive relevant details based on task complexity and urgency; minimizing distractions and enhancing situational awareness~\cite{bell2001view}. For example, minimal guidance is provided during routine navigation, while more detailed cues are offered during complex tasks like landing in adverse weather~\cite{duchowski2018index}. 

\vspace{-0.05in}
\myparagraph{B3. Real-Time Modality Switching}
A neuroadaptive system should provide real-time switching between various (e.g., visual, auditory, and textual) modalities. Given a user's processing bandwidth, the switching should be based on their workload. Different modalities have varying response times, cognitive load, and attention demands, making adaptive switching crucial for optimizing pilot performance in real-time~\cite{kasarskis2001comparison}.

\vspace{-0.05in}
\myparagraph{B4. Personalization Based on Pilot Expertise}
The system should account for individual differences in pilot experience. Novice pilots may require detailed guidance, while experts often benefit from less intrusive support \cite{kasarskis2001comparison, wickens2008multiple}. This differentiation enhances performance and ensures scalability across varying experience levels.

\vspace{-0.08in}
\section{Formative Study}
\revision{In the formative study, we explore how a system can employ \textbf{adaptive rules} to dynamically adjust interaction modalities and feedback, thereby maintaining pilot performance. At the end of this study, we extend and translate the findings into actionable design requirements for an adaptive system. To begin, we explore the following research questions:}
\begin{enumerate}[start=1,label={\bfseries RQ\arabic*},left=0.1cm]

\vspace{-0.08in}

    \item \label{rq1} How do varying cognitive states impact a pilot’s ability to process and comprehend information during flight operations?

\vspace{-0.1in}

    \item \label{rq2} In what ways could a dynamic guidance system enhance performance and safety?

\vspace{-0.1in}

    \item \label{rq3} How do different information modalities (visual, auditory, text) affect cognitive states, and how could adaptive modality switching optimize task execution for pilots?
\end{enumerate}

\vspace{-0.08in}
\subsection{Expert Interview}
We recruit three licensed, experienced pilots as experts and interview them to glean insights for the RQs. Each expert is experienced with flight training systems, multimodal feedback design, and adaptive flight support systems. 

\vspace{-0.05in}
\myparagraph{Expert Background}
\revision{\textbf{E1} is a retired fighter pilot with over 10 years of active service (Male, Age=54). E1 works at a pilot retirement office, contributing to the design of training systems for novice pilots.~\textbf{E2} is an active flight instructor with over 15 years of flight experience, specializing in human factors and cockpit interactions (Male, Age=48).~\textbf{E3} is a researcher in aviation and flight technology with extensive experience in flight simulations and multimodal feedback systems, currently working in research and development for a major airline's flight training programs (Female, Age=39).}

\vspace{-0.05in}
\myparagraph{Method}
We conducted semi-structured interviews  with each expert \revision{(see supplementary section ''Formative Study Questions'')}. We presented several initial adaptive design rules based on \textbf{B1-4} and gathered their opinions on feasibility and improvements. All interview recordings were transcribed and qualitatively coded. \revision{Three researchers independently performed open coding on the transcriptions and reached a consensus through collaborative discussions. }

\vspace{-0.08in}
\subsection{Formative Study Results}

\vspace{-0.05in}
\myparagraph{Importance of Modality Switching Between Stages}
Experts emphasized that during takeoff, experienced pilots—familiar with the essential actions—may find repeated visual cues redundant or distracting; audio prompts often suffice. However, if a pilot overlooks a crucial action, the system should switch to provide both audio and visual reminders. E1 noted, \participantquote{For key actions like retracting the landing gear at 50 meters altitude, if forgotten, an integrated audio-visual prompt would help correct the oversight.} This highlights the need for context-aware systems that adapt feedback based on pilot engagement and task familiarity.

\vspace{-0.05in}
\myparagraph{Reducing Error Rates with Multimodal Feedback}
E2 mentioned that pilots sometimes make critical errors, such as incorrectly setting descent rates. An adaptive system providing immediate auditory prompts can report the error and provide correct operational instructions. E3 added that real-time visual and auditory prompts have great potential for minimizing errors during high-stress tasks like takeoff. E3 stated, \textit{"The system’s ability to correct errors with multimodal cues—particularly a combination of visual and auditory feedback—reinforces proper procedures and significantly reduces the likelihood of mistakes."}

\vspace{-0.05in}
\myparagraph{Managing Fatigue and Boredom with Varying Information Levels}
E2 noted that reliance on electronic displays during taxiing can lead to disengagement due to task repetition. Providing dynamic information—such as alerts about weather conditions, flight path deviations, or upcoming obstacles—could help maintain attention and reduce fatigue. E3 suggested incorporating text and visual information alongside audio to offer anticipatory warnings, especially for less experienced pilots. E2 added, ``By incorporating visual and textual information into routine tasks, the system can help prevent fatigue and better prepare pilots for unexpected events.''

\vspace{-0.05in}
\myparagraph{Multimodal Feedback as Safety Assistant}
All experts agreed that multimodal feedback systems should function as safety assistants rather than rigid command structures. E1 emphasized the need for flexibility, noting that pilots vary in flying styles and endurance; thus, the system should not impose a one-size-fits-all approach. E2 supported this view, highlighting that adaptive feedback should be perceived as a helpful guide. 

\vspace{-0.08in}
\subsection{Design Requirements} 

\vspace{-0.05in}
Based on the above findings, we summarize four design requirements for developing an adaptive multi-modal Feedback system.
\begin{enumerate}[start=1,label={[\bfseries R\arabic*]}]

\vspace{-0.08in}
\revision{\item \label{dr:overload} \textbf{Prevent Information Overload and Redundancy Through Context-Aware Modality Switching.}
The system should tailor feedback type and frequency using concise prompts for routine tasks, and escalating to integrated audio-visual cues if critical actions are missed. This approach avoids unnecessary repetition and ensures essential information is always highlighted.}

\vspace{-0.08in}
\revision{\item \label{dr:dynamic} \textbf{Dynamically Adapt Information Load to Manage Task Complexity and Reduce Boredom.} The system should vary both the feedback modality and level of detail according to task demands, pilot experience, and potential fatigue. For instance, add contextual visual or textual details during taxiing, while prioritizing concise audio prompts during takeoff or approach.}

\vspace{-0.05in}
\revision{\item \label{dr:error} \textbf{Provide Timely, Contextual Multimodal Feedback to Address Pilot Errors as a Flexible Safety Assistant.} 
The system should detect pilot errors and immediately supply corrective cues tailored to the specific flight phase. For example, if a pilot forgets to retract landing gear at the prescribed altitude, audio and visual alerts should be triggered to highlight the oversight and reinforce correct procedure.}

\end{enumerate}

\vspace{-0.1in}
\section{\systemname system}

\systemname provides adaptive guidance through text, audio, and interactive cues based on real-time assessments of the pilot’s cognitive load and performance. The system dynamically adjusts the combination and quantity of these modalities depending on whether the cognitive load is sub-optimal (either an overload state or an underload state). This section outlines the design decisions behind our system, identifies the challenges they pose, and describes how we address these challenges.

\subsection{Adaptive strategies}
\label{sec:adapt_strategies}
\revision{In alignment with all design requirements(~\ref{dr:overload}~,~\ref{dr:dynamic}~,~and~\ref{dr:error}), we use 28 adaptive strategies(see supplementary section ``Adaptive Rules''). We use cognitive workload as a way to estimate user disengagement and information overload.} These strategies are based on the following three key cognitive workload theories:

\vspace{-0.05in}
\myparagraph{Underload Scenarios} Often times this is due to a lack of focus~\cite{sweller1988cognitive}. The system should provide more information and use diverse modalities (visual, audio, text) to stimulate engagement and prevent errors~\cite{zeng2024visual}. The system should also increase detailed information. As cognitive demand is low, introducing additional complexity may help prevent errors by minimizing the misallocation of attention. 

\vspace{-0.05in}
\myparagraph{Overload Scenarios} During a cognitive overload, the system uses mainly the visual cues to keep pilots calm and focused~\cite{spence1997audiovisual}. Here, visual cues only contains information for the essential tasks, providing the most critical information to guide pilots through decision-making processes~\cite{mousavi1995reducing}.

\vspace{-0.05in}
\myparagraph{Optimal Scenarios} When the cognitive load is optimal, the system employs a mix of visual and audio prompts, providing information to assist the pilot while avoiding unnecessary distractions~\cite{parasuraman2008mitigating}. The system aims to deliver just-in-time feedback, giving enough information to assist the pilot without causing distraction. 




\vspace{-0.08in}
\subsection{Multimodal Feedback}
\revision{To address \ref{dr:overload}~and~\ref{dr:error}}, \systemname's selection of the modality ensures that the pilot receives information in a minimally disruptive manner. The modalities available for instruction delivery include visual, audio, text, or combination of these. Textual cues are presented via a pop up window in the pilots field of view. Visual cues are presented with a ghost hand pointing to the current action the pilot must perform. Audible cues are played via the headset's speaker to the pilot. \revision{To provide error correction and address expertise differences, information is displayed at an adaptive rate. If a pilot has not performed an action in 20 seconds or performs an error, information is sent to the user about the current step.}

\revision{
\myparagraph{Modality-Selection Logic}
Our system relies on a set of adaptive strategies~\ref{sec:adapt_strategies} to link a cognitive state (underload, optimal, overload) with modalities. For example, in \textit{overload} scenarios, the system chooses either “visual only” or “visual + minimal audio” to avoid further mental strain from concurrent text or extended voice prompts. In contrast, underload conditions trigger “visual + audio + text” or “visual + text” to foster deeper engagement. 
}


\vspace{-0.08in}
\subsection{Adaptive Information Load}
To address~\ref{dr:dynamic}, \systemname~calibrates the information load (e.g., length and details in text or duration of audio) based on the pilot’s cognitive state to avoid overwhelming them while ensuring that they remain engaged and focused. If the pilot is experiencing overload, the system limits the information load to essential commands only, minimizing the risk of further cognitive strain. In an optimal cognitive state, the system balances the information load by providing both command instructions and additional context, such as flight information or flight route details, which enhances the pilot’s understanding and situational awareness. When the pilot is underloaded, the system increases the information load by including comprehensive details, such as environmental updates, to ensure that the pilot remains engaged and maintains a high level of situational awareness. This dynamic adjustment of information load helps the pilot to remain focused and efficient without becoming overwhelmed or disengaged.

\vspace{-0.1in}
\section{\systemname Implementation}

\subsection{Design Decisions}

\begin{figure}
    \centering
    \includegraphics[width=\linewidth]{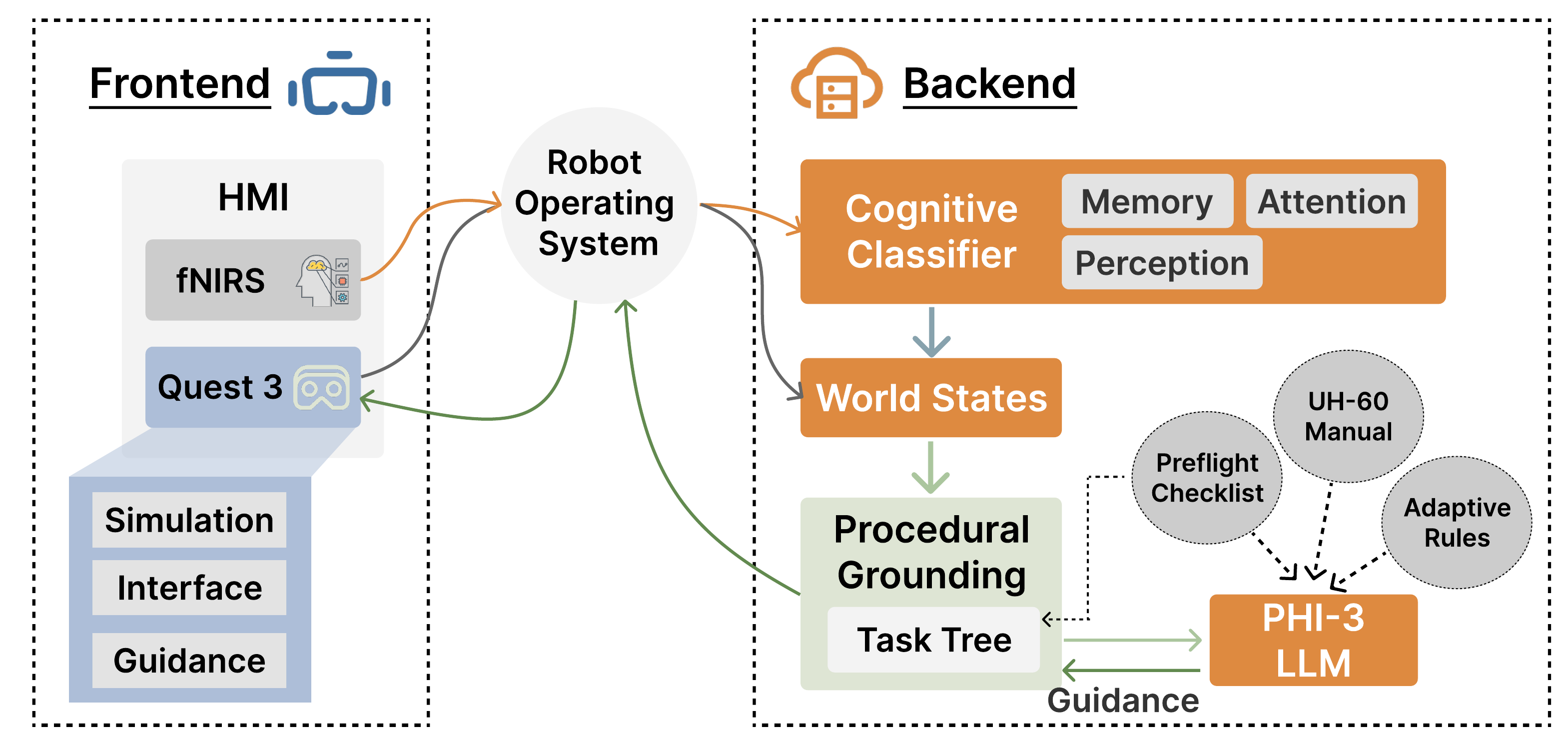}
    \caption{The front and backend systems. A change in arrow color represents a change in information.}
    \label{fig:system_flow}

\vspace{-0.2in}
\end{figure}


\vspace{-0.05in}
\myparagraph{Apparatus} \systemname is divided into a front-end VR system built in Unity and a back-end system in Python, allowing the VR system to run on the headset while complex calculations are handled remotely or on a separate server (see Fig.~\ref{fig:teaser} (C-D)). \revision{Audible instructions are generated using WindowsVoice. We use a Biopac 2000s fNIRS system to detect and monitor participants' workload states in real-time. The experiment is conducted using the Quest 3 VR headset.}

\vspace{-0.05in}
\myparagraph{VR Simulator for UH-60 Black Hawk Cockpit}
Our test bed builds on a high-fidelity UH-60 VR simulation, offering a flexible platform to adjust cockpit events and present information to the pilot \cite{guthridge2023evaluating}. We implement a section of a preflight checklist of a UH-60. We use the preflight checklist for our scenario as it is easily implementable, structured, and prone to mistakes due to complacence and repetitiveness \cite{degani_human_1991}. To replicate the real-world pre-flight checklist process, we added an interactive recreation of the paper checklist.
\revision{The full preflight checklist can be found in section ``Preflight Paper Checklist'' of supplementary materials.}

\vspace{-0.05in}
\myparagraph{Reasoning Model Selection}
\systemname uses a quantized version of Microsoft's PHI-3 multi-modal LLM \cite{kaplan2020scaling}, selected for its reasoning and question-answering capabilities \cite{kenton2019bert}. Quantization improves inference speed and allows hosting on a local GPU.

\vspace{-0.05in}
\myparagraph{Streaming Data}
Data between the front and back end is streamed through the Robot Operating System (ROS) architecture \cite{koubaa2017robot}, enabling modular functionality. PHI-3 runs on a GPU cluster, other back-end nodes on a separate computer, and the front end on the headset. ROS BAGs record all system data with normalized timestamps for replayability and data extraction. However, this design can sometimes lead to disconnects between user actions and system state. To address this, we provide a dynamic \emph{error dashboard} that audibly alerts pilots when an error occurs. Pilots are instructed to consult the dashboard only after hearing the alert and then return to the paper checklist. While this may add cognitive load, it ensures accurate action tracking. \revision{During data collection this contributed to an approximate 0.053 +- 0.021 error rate. A full distribution of these false errors can be found in the supplementary material section ``False Error Rates''. Note these errors are not included as errors in our quantitative evaluation.}



\vspace{-0.05in}
\myparagraph{Functional Near-Infrared Spectroscopy}
Cognitive workload detection employs fNIRS, a non-invasive technology that measures brain activity through changes in blood oxygenation. The fNIRS sensors are positioned to capture brain activity in the prefrontal cortex, which is linked to executive cognitive functions. The device emits two wavelengths of near-infrared light into the scalp and records their absorption by the underlying tissue to detect relative changes in oxygenated and deoxygenated hemoglobin.  These measurements inform our workload classifier to predict a pilots given workload state across working memory, attention, and perception. 
\revision{Fig.~\ref{fig:workload_flow} shows our real-time classifier for workload. The underlying models for classification (Sec.~\ref{sec:pcsm}) are multinomial regression models that take as input preprocessed fNIRS readings across the prefrontal cortex.

\vspace{-0.05in}
\myparagraph{fNIRS Prepossessing}
fNIRS data is captured at a rate of 10 Hz using a 18 channel Biopac Imager 2000S. Data preprocessing is done over a sliding window of 10 second chunks using the same methods described in \cite{mckendrick_theories_2019}. The first motion artifacts are corrected using the Daubechies (db5) wavelet filter, and the wavelet coefficients are computed using a threshold of 0.1 \cite{molavi_wavelet-based_2012}. High-frequency noise, respiration, and cardiac cycle effects are removed through a 0.12 Hz low-pass filter \cite{ayaz_sliding-window_2010}. Optical density is calculated and base lined using the log of the filtered signal divided by the average. Oxygenated (HbO) and Deoxygenated (HbR) are base lined and calculated through the modified Beer-Lambert law \cite{ayaz_using_2012a, ayaz_optical_2012b}.

\vspace{-0.05in}
\myparagraph{fNIRS Baseline Calibration} 
The baseline calibration task used for fNIRS imaging uses the same method as described in \cite{mckendrick_theories_2019}. The purpose of the baseline task is to elicit consistent brain activity to control for variability between participants. For details of the baseline fixation task, please refer to ``fNIRS Baseline Task'' in the supplementary materials. }

\vspace{-0.05in}
\revision{
\myparagraph{System Flow}
Figure \ref{fig:system_flow}  provides a block diagram overview of the data flow between the VR headset, the Python-ROS back end, and the large language model PHI-3. The system tracks world states at 10 Hz to query cockpit parameters, previous user behavior, and cognitive state changes (attention, working memory, perception). Procedural grounding is tracked through task trees created from the procedure checklist which tracks progress and errors as world states change. Information from the world states and procedural grounding components inform the PHI-3 reasoning model, which selects a suitable adaptive strategy.}

\vspace{-0.08in}
\subsection{Models for Adaptation}
\subsubsection{Pilot's Cognitive State Modeling}
\label{sec:pcsm}

\begin{figure}
    \centering
    \includegraphics[width=\linewidth]{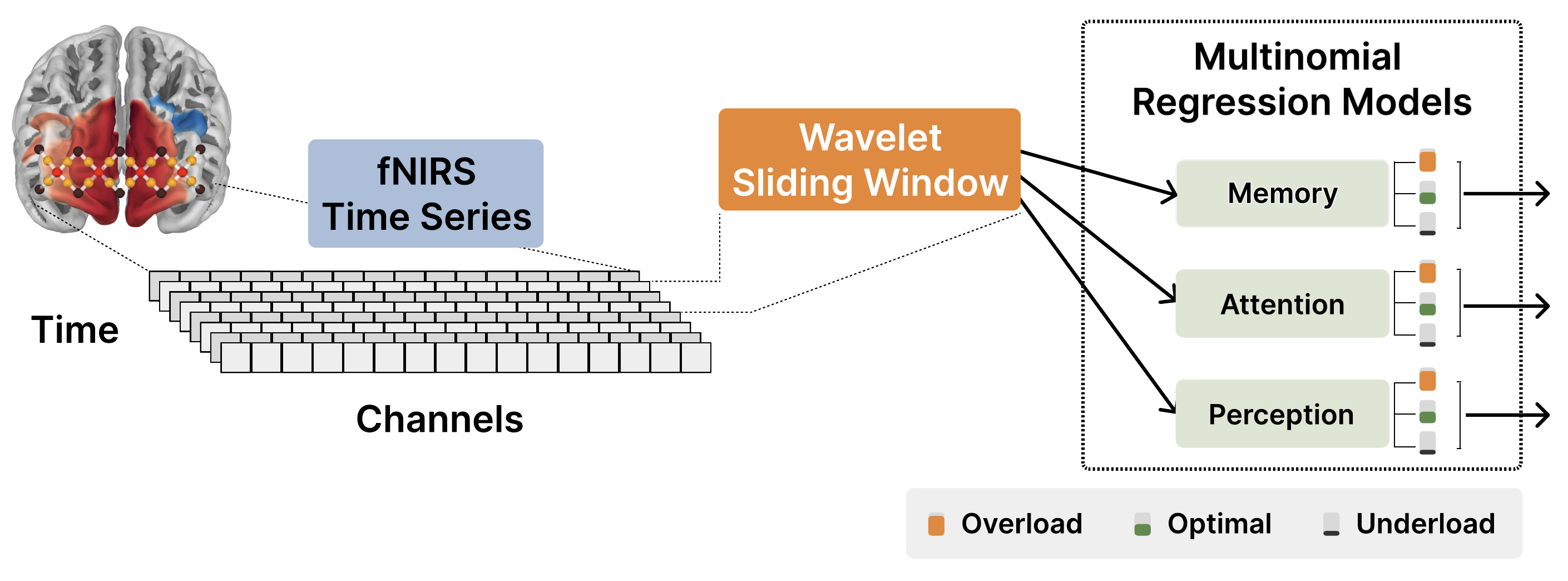}
    \caption{Cognitive classifier: fNIRS data is preprocessed using wavelet filters over a sliding window then the 3 cognitive facets are classified.}
    \label{fig:workload_flow}

\vspace{-0.2in}
\end{figure}

The system predicts the pilot's cognitive load state using fNIRS sensor input, see Figure \ref{fig:workload_flow}. \revision{The processed input contains HbO and HbR concentrations and is then classified into three facets using Mckendrick's method~\cite{mckendrick_theories_2019}: perception, attention, and working memory. Our classifiers rely on multinomial symbolic regression, previously trained on a Rasch model labeling methodology, updating at 10hz. The classifier will output three discrete labels as: "underload," "overload," and "optimal." Full documentation of cognitive classifier architecture and training data can be found in \cite{mckendrick_theories_2019}.}

\revision{The term ``underload'' is labeled when fNIRS signals indicate a lower-than-expected resource utilization, potentially risking inattention. ``Overload'' arises when signals indicate high resource utilization, risking errors from excessive mental demands. ``Optimal'' reflects moderate, task-appropriate resource utilization, where pilots demonstrate stable facet utilization.}


It is important to note that these workload states implicitly account for an individual's cognitive capacity via modeling cognitive capacity during the state labeling process of training the classifiers. They are not equivalent to measures of high, medium, or low workload. Each cognitive facet uses a separate classifier, operating simultaneously during recording sessions. These classifiers have shown strong performance across tasks and between participants \cite{mckendrick_theories_2019}. \revision{See \cite{mckendrick_theories_2019} for full documentation of our cognitive models and training data.}

\vspace{-0.08in}
\subsubsection{PHI-3 Prompt Engineering}
The PHI-3 LLM is fine-tuned to support the \systemname adaptive strategies. We use its reasoning capacity to provide adaptive instructions to novice flight trainees, ensuring that the guidance aligns with the pilot's current cognitive load, focus, and procedural context. It's reasoning prompt structure is divided into three components: instructions, few-shot examples, and real-time context. The instructions laid out the context for PHI-3 around the UH-60 cockpit from a tokenized UH-60M operation manual. The instructions also contain information regarding cognitive workload definitions and define all adaptive strategies, see the supplementary materials 6 for full list of instructions. The few-shot examples provide contextual examples of how to use the adaptive strategies in context to adapt guidance. Finally, real-time context is given to the PHI-3 model during runtime. The real-time context includes: 1) cognitive workload across all 3 facets over the last 10 seconds; and 2) the pilot's current tasks; and 3) the pilot's current gaze focus.
\revision{The full example prompt is provided in section ``Prompt Example'' of supplementary materials.
}

%



\myparagraph{Design for Reasoning}
We use a chain-of-thought reasoning that uses the intermediate steps to generate the appropriate output. For instance, when the pilot experiences high cognitive overload with optimal attention, the chain of thought involves assessing how this cognitive state might affect the pilot’s ability to execute the next steps accurately. The model is instructed to present this reasoning as part of its output.

\vspace{-0.05in}
\myparagraph{Output}
The final output in each example includes the selected modality and the specific instructional load, tailored to the pilot's current cognitive workload state and task instructions. The instructions focus on reinforcing the next steps in a manner consistent with the pilot’s cognitive capacity. The model provided reasoning for its choice of modality type and information load levels based on the cognitive workload levels, behavioral data, and task information. It also specifies the modality type to use and the information to present to the pilot, particularly for audio and text cues.

\vspace{-0.08in}
\section{Evaluation: VR-Simulated Flight Task}

To evaluate the effectiveness of \systemname, we conducted an empirical experiment involving eight pilots with varying levels of expertise to understand their task performances using \systemname against non-adaptive and randomly adaptive conditions. \revision{We aim to further elicit potential benefits of using automated multimodal feedback from generative AI in cockpit operations.}
\revision{We use a mixed design for this study, with a within-subject component focusing on quantitative evaluation of pilots' task performance over three guidance conditions (baseline, random, adaptive). We also collect the pilots' expertise level as a between-subject variable in the mixed design to qualitatively interview them for their feedback. As a result, the \textit{quantitative evaluation} of this experiment will focus on the within-subject task performance and \textit{qualitative evaluation} will focus on between-subject component to understand the nuances between expertise levels and their responses to an automatic assistance.}

\vspace{-0.05in}
\myparagraph{Conditions}
This study uses three conditions: baseline, random, and adaptive see supplementary materials 1. In the baseline condition, pilots had no virtual guidance besides the error correction and could only walk through the checklist. In the random guidance presentation, pilots were presented the three modalities at random, with the information load of the text and audio modalities being a hard coded action instruction and visual description. In the adaptive guidance condition, pilots were presented with the modality type guidance based on the output of the PHI-3 LLM, with the information load being an adapted version of the same hard coded action instruction and visual description as the random condition. Conditions were randomized via a Latin squares randomization for each participant. In both the random and adaptive conditions, guidance was presented at a fixed rate of 10 seconds after a previous correct action was performed, and if no new correct action had been initiated.

\vspace{-0.08in}
\subsection{Participants}
\revision{Due to the challenges in recruiting licensed pilots, not every individual could attend the quantitative evaluation; we describe the recruiting details below.}

\vspace{-0.05in}
\myparagraph{Participants for quantitative evaluation}
We recruited eight participants for the quantitative evaluation. (4 males, 4 females) with an age range from 23 to 61 years (M = 40, SD = 12.7). These pilots come from a mixed background of recreational fixed-wing (5), fighter pilots and flight instructors (2), and Black Hawk pilot (1). Due to technical performance issues, some fNIRS data sessions from two participants could not be used in our final quantitative evaluation.

\revision{Their familiarity with VR was distributed as follows: four reported no prior VR experience, two indicated annual or occasional VR use, and two reported higher familiarity.}

\vspace{-0.05in}
\myparagraph{Participants for qualitative evaluation}
\revision{To have more comprehensive, diversified responses from pilots with different backgrounds, we invited an \textbf{additional five more pilots}(three novice pilots, one additional Black Hawk pilot, and one fixed-wing instructor) for the interviews. They used the \systemname to respond to qualitative interviews. In total, 13 participants completed the interviews.}

\vspace{-0.05in}
\myparagraph{Black Hawk Expert Background} Both professional pilots (denoted as \textbf{E1} and \textbf{E2}) have extensive experience with actual Black Hawk flying experience. \textbf{E1} is an expert in aviation research with over 10 years of experience flying Black Hawk helicopters. They are experienced in advanced piloting techniques and the integration of technology into cockpit operations. \textbf{E2} is a retired military veteran with over a decade of service, including significant operational experience with Black Hawk helicopters. They have been involved in various high-stress, mission-critical scenarios.

\vspace{-0.1in}
\subsection{Procedure}
Before the experiment, both the paper checklist and a demonstration video of the preflight procedures were sent to the participants, allowing them to review the material in advance. First participants were given a 10-minute introduction to the UH-60 preflight interface, covering all relevant controls. 

The pilot was instrumented with the fNIRS and VR headset, baselined while wearing the headset, and put into the VR cockpit see supplementary materials 1. The pilot was informed on how to interact with the checklist, the cockpit systems, and shown the error correction dashboard. While in the baseline condition the pilot was instructed to only use the checklist unless they hear an audible error noise, and then to use the error dashboard to correct their error before returning to the checklist. We provided printed copies of the paper checklist and physical representations of the operational panels. Before starting the formal experiment, participants were given a walk-through of specific interactions they would do in the cockpit and spent approximately 5 minutes acclimatizing to the VR environment.

\revision{Since each participant will complete all three guidance conditions, we collected repeated measurements across multiple procedures (9 steps each) per pilot. This repeated-measures approach, combined with our mixed effects models, increases statistical power despite a small number of participants. However, given the niche population and sample size, caution is advised in generalizing the results.}

During each trial, pilots performed a 9-procedure UH-60M pre-flight checklist. If a pilot became stuck on a procedure, the experimenter would verbally help the pilot get back on track with the checklist.
Workload, errors, and guidance tips were recorded during the trials. After all 3 conditions were completed the pilot was asked a series of semi-structured interview questions on specific feedback related to the pilots runs and background.

\revision{We gather qualitative feedback through semi-structured interviews. We ask a set of predefined questions on areas such as usability, the effectiveness of different modalities, and perceived workload levels (Full list of questions can be found in supplementary materials section ''Qualitative Evaluation Questions''). Participants were then invited to freely share additional thoughts and suggestions without any prompts.}
Upon completing the three conditions, participants were asked these interview questions. Each participant received a \$100 compensation. The experiment lasted approximately 2 hours.




\vspace{-0.08in}
\subsection{Quantitative Evaluation}

\begin{figure}
    \centering
    \includegraphics[width=0.95\linewidth]{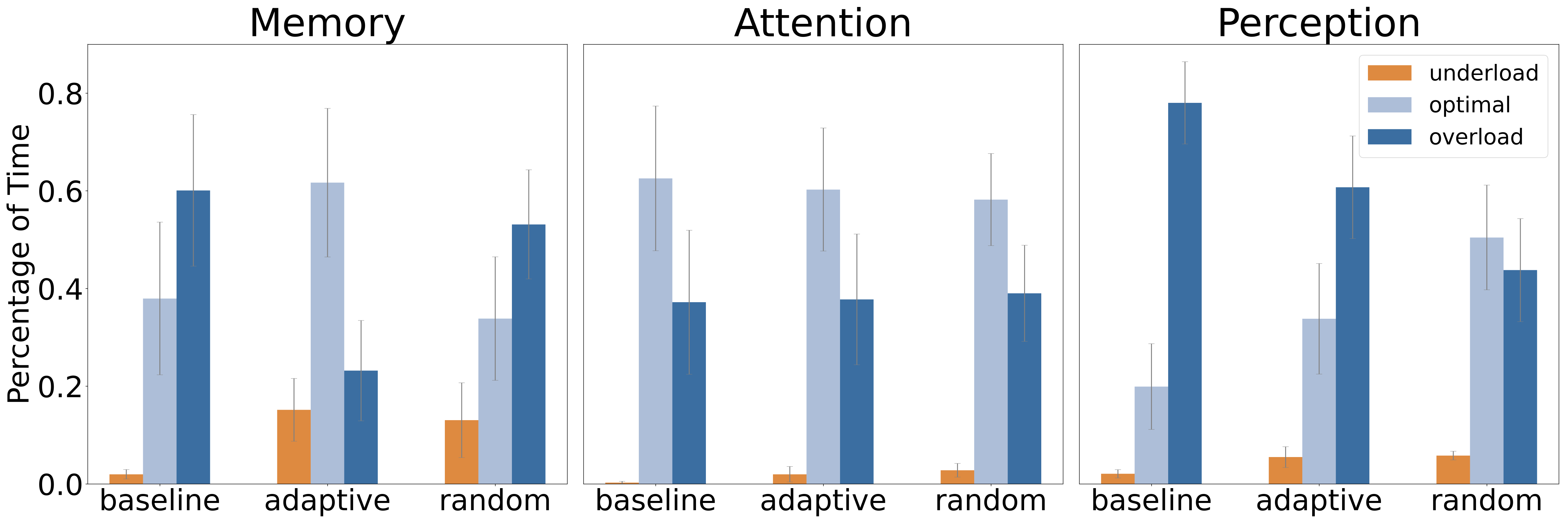}
    \caption{All Workload Across Conditions}
    \label{fig:all_workload}
    \vspace{-.15cm} 
\end{figure}

Fig.~\ref{fig:all_workload} shows an overview of workload states in conditions and facets. 

\begin{figure}
    \centering
    \includegraphics[width=0.95\linewidth]{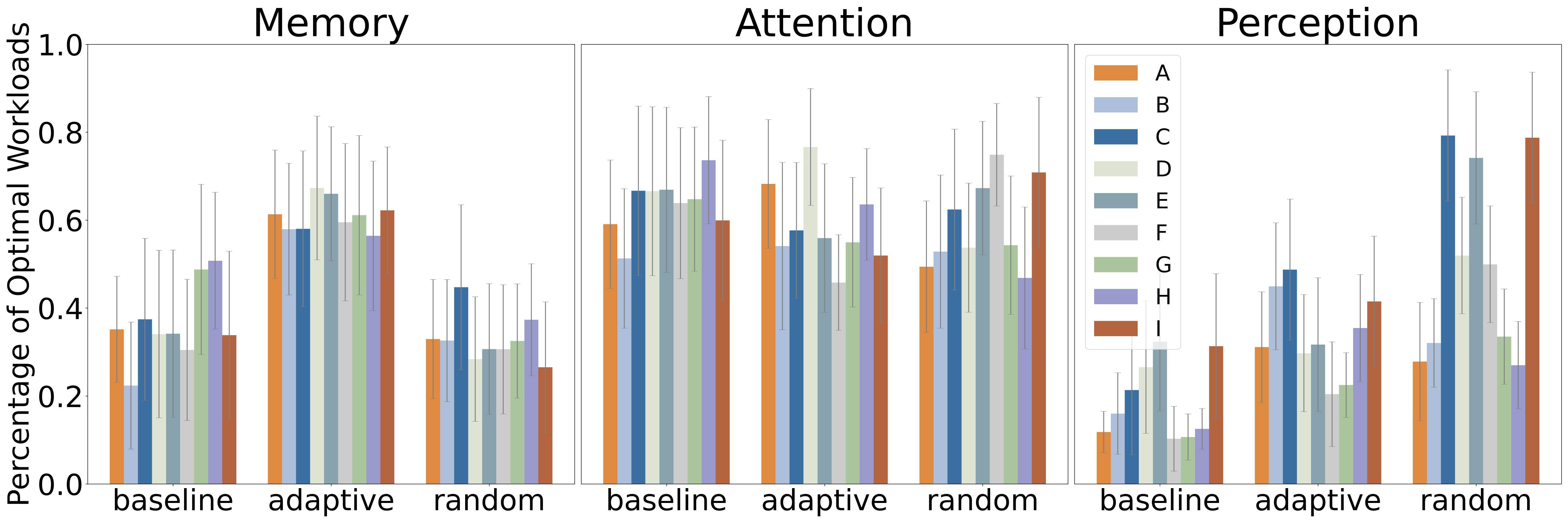}
    \caption{Optimal Workload Across Conditions and Procedures. We see a optimal increase in the adaptive condition for memory. A-I corresponds to a procedure.}
    \label{fig:optimal_workload}
    \vspace{-0.2in} 
    
\end{figure}

\textbf{Working Memory} was analyzed as a logistic mixed effects model with a binomial encoding of optimal(1) vs non optimal(0) working memory rates. We see significantly different rates of optimal memory states both in baseline vs adaptive \((\text{beta}=-0.685; \text{se}=0.042; \text{z}=-16.173; \text{p}<0.001)\) and random vs adaptive \((\text{beta}=-1.391; \text{se}=0.045; \text{z}=-30.737; \text{p}<0.001)\). The log-odds ratios indicate a larger optimal workload rate in the adaptive condition over both random and baseline. We see in Fig.~\ref{fig:optimal_workload} a trend in optimal working memory while using the adaptive guidance condition. For all procedures, we see an increase in optimal working memory states in contrast to both baseline and random conditions.



\textbf{Attention} was analyzed as a  logistic mixed effects model with binomial encoding for optimal(1) vs non optimal(0) attention workload states. No significant differences were observed for optimal load rates both in baseline vs adaptive \((\text{beta}=0.004; \text{se}=0.034; \text{z}=0.124; \text{p}=0.901)\) and random vs adaptive \((\text{beta}=0.023; \text{se}=0.036; \text{z}=0.630; \text{p}=0.529)\). We further see in Fig.~\ref{fig:optimal_workload} no visible change between conditions and procedures. 
\begin{figure}
    \centering
    \includegraphics[width=0.95\linewidth]{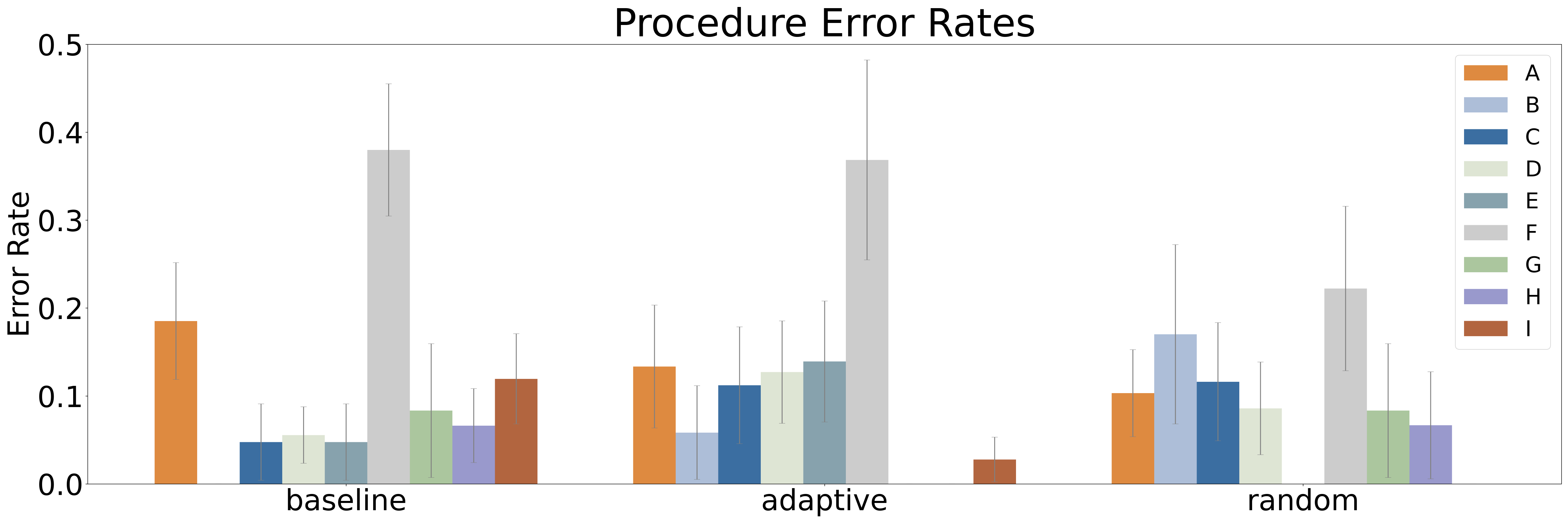}
    \caption{Error Rates Across Conditions and Procedures. We see a slight decrease in error rates in the baseline. A-I corresponds to a procedure.}
    \label{fig:error_rates}
    \vspace{-0.2in} 
\end{figure}

\begin{figure}
    \centering
    \includegraphics[width=0.95\linewidth]{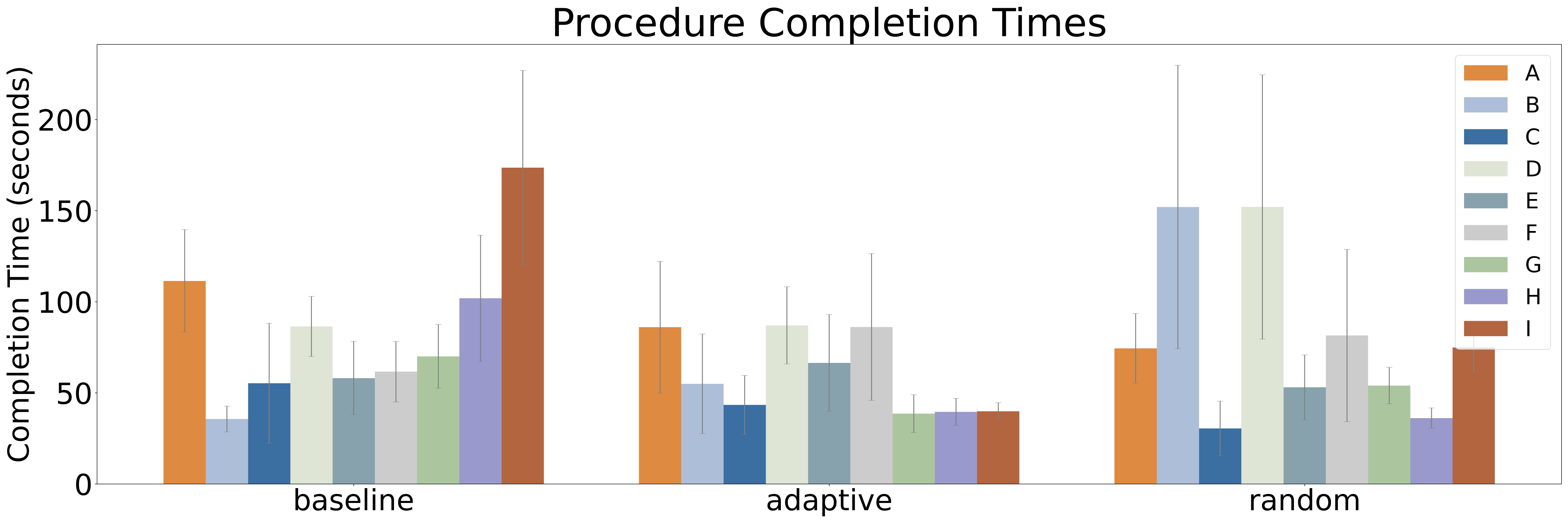}
    \caption{Completion Time Across Conditions and Procedures. We see a decrease in completion time in adaptive. A-I corresponds to a procedure.}
    \label{fig:completion_time}
    \vspace{-0.2in}  
\end{figure}


\textbf{Perception} was analyzed as a logistic mixed effects regression with binomial encoding of optimal(1) vs non optimal(0) perception states. We see significant differences in both baseline vs adaptive \((\text{beta}=-1.403; \text{se}=0.038; \text{z}=-36.481; \text{p}<0.001)\) and random vs adaptive \((\text{beta}=0.421; \text{se}=0.036; \text{z}=11.640; \text{p}<0.001)\). The log-odds ratios suggest an improvement in optimal perception states over baseline for both random and adaptive with a larger overall improvement with the random condition. We see in Fig.~\ref{fig:optimal_workload} a visible trend of optimal states improving between baseline and both guidance conditions. When we look at the adaptive vs random conditions we see that in some procedures optimal workload is improved with the adaptive condition and some with the random condition. A further dive into the procedures that improve with adaptive vs random; we can see that procedures where the pilot inputs data, show increased optimal workload states during the adaptive condition. Whereas, procedures where the pilot must perceive and confirm information show increased optimal states in the random condition.


\textbf{Error Counts} were analyzed as a generalized mixed effects poisson model  on both adaptive vs baseline and adaptive vs random conditions. We calculate error rate ratios for adaptive vs baseline of \((\text{rate ratio}=0.644; \text{~z}=-2.276; \text{p}=0.0228)\) and adaptive vs random of \((\text{rate ratio}=0.962; \text{~z}=-0.222; \text{p}=0.8244)\). We observed a significant difference  in the adaptive vs baseline comparison, indicating a lower error count with baseline relative to adaptive. A potential cause of this could be from a loss of focus and complacency with the guidance tips as opposed to the higher concentration required for the baseline. We can further examine the error rates across procedures and conditions in, Fig.~\ref{fig:error_rates}. The visual results show a similar error rate across all conditions and procedures

\textbf{Completion Time} was analyzed as generalized mixed effects gamma model over the total completion time(seconds). No significant differences were observed in the time to complete the 9 procedures for both baseline vs adaptive \((\text{beta}=0.342; \text{se}=0.333; \text{z}=1.029; \text{p}=0.3034)\) and random vs adaptive \(( \text{beta}=0.280; \text{se}=0.333; \text{z}=0.842; \text{p}=0.3998)\). Although the results are not significant the positive standardized beta coefficients indicate trend toward a lower time to overall completion in the adaptive condition vs both random and baseline. Suggesting that participants had an easier time going through procedures with the adaptive guidance. Fig.~\ref{fig:completion_time}, shows the completion time in seconds between procedures and conditions. We can see a general trend across all procedures that the adaptive condition has a faster completion time.

\vspace{-0.08in}
\subsection{Qualitative Evaluation}
To better understand the effectiveness and first-hand experience of using \systemname{}, we conducted semi-structured interviews with two professional Black Hawk pilots and 11 licensed pilots. 
We will structure the following sections feedback as follows: pilots who were novices with the system will be marked as P(1-8); fighter pilots FP(1-2); flight instructor I(1); and expert E(1-2).
%
The interviews lasted approximately 30 minutes each and focused on the experts’ experiences using the adaptive system, their assessment of the system’s usability in real-world scenarios, and potential improvements. 

\vspace{-0.08in}
\subsubsection{Interviews}

\vspace{-0.05in}
\myparagraph{Tailoring Information to Preflight Task Demands}
The \systemname{}’s ability to provide relevant and task-specific information during preflight operations was a strength highlighted by some participants. This feature was particularly useful for complex preflight tasks, such as setting waypoints, configuring flight plans, or verifying control settings.
\textit{P5} mentioned, \participantquote{When I was setting waypoints, the system knew exactly when to give me the next step, without bombarding me with irrelevant details. 
It was like it knew what I needed right at that moment.} 
Moreover, the \systemname{}’s ability to adapt the amount of information based on the complexity of the preflight task was noted. 
\textit{P6} pointed out how the adaptive system provided more detailed auditory and textual guidance when the task required higher attention: \participantquote{During preflight checks, the system would switch to more detailed visual feedback whenever I needed to double-check something, but it wouldn’t overwhelm me with too much at once. That really helped me stay on track.}
In contrast, some participants who used the baseline versions felt that they lacked the same level of task-specific adaptation.
\textit{P5} reflected, \participantquote{With the baseline, I often wasn’t sure if I was doing things correctly because it didn’t adjust to the pace of the task. I had to constantly check things myself, which slowed me down.}

\vspace{-0.05in}
\myparagraph{Seamless Modality Transitions Can Enhance Efficiency}
Participants found that the transitions between visual, auditory, and text-based feedback were effective. They appreciated how the adaptive system would switch between modalities based on the task context. 
\textit{P6} explained, \participantquote{When I was looking for a specific button, the system quickly shifted to visual cues, showing me exactly what to do. The transition felt natural, and I didn’t have to pause or adjust.} 

However, participants emphasized that this fluid modality switching worked best when it was applied precisely and only when necessary. \textit{P5} noted, \participantquote{The system never overloaded me with information—when I needed a quick visual, it switched smoothly, and when a task didn’t need a lot of input, it didn’t give me anything extra.} This balance between providing sufficient guidance and with cognitive states seemed to be an advantage of \systemname{}.

\vspace{-0.05in}
\myparagraph{Error Detection and Correction Support}
Participants noted the adaptive system’s ability to provide additional information to guide them in detecting and correcting mistakes. Unlike the baseline system, which lacked dynamic feedback, both the random and \systemname{} provided cues that helped users recover from errors without needing to manually cross-check steps.
\textit{P6} highlighted how the adaptive system, in particular, simplified error correction: \participantquote{If I made a mistake, the tip window immediately pointed out what I should do next, and I didn’t need to look back at the dashboard or compare it with the paper checklist. One time, I couldn’t figure out where I went wrong, but the adaptive system used the green hand to show me exactly what to do next—it was incredibly helpful.} This immediate, step-by-step guidance minimized downtime, allowing participants to quickly recover from errors and continue the task.

However, some participants suggested room for improvement. \textit{P8} remarked that while the random version offered more background information, the adaptive version was more focused on the immediate next step. \participantquote{The adaptive system told me what to do next, but when I made a mistake, I wasn’t sure which step I needed to go back to.} 
This feedback highlights the need for a more comprehensive error recovery feature, where the \systemname could guide users back to the specific point of error, helping them understand their mistakes and enabling a smoother, more intuitive correction process.

\vspace{-0.05in}
\myparagraph{Adaptive Guidance Potential in Training}
The experts provided an overall positive assessment of the adaptive system, particularly highlighting its suitability for training applications. 
Both experts noted that while experienced Black Hawk pilots would not use this system during regular operations, it holds significant potential in training scenarios, especially for novice pilots who are still familiarizing themselves with cockpit controls and procedures.

\textit{E1} commented that for experienced pilots who already know the cockpit layout and procedures, the adaptive system would not be necessary during routine operations.
However,they emphasized that \participantquote{In a training scenario, especially for pilots still learning the locations and functions of various controls, the adaptive system could be highly beneficial.} This aligns with feedback from other participants, such as \textit{P4}, who valued the system for its ability to assist with procedural familiarity and visualization. \textit{P4} stated that the adaptive system was particularly helpful in replacing traditional methods like "chair flying," where pilots mentally simulate flight tasks. 
\participantquote{Instead of having to imagine everything, the system presents it for you, which helps you become more familiar with the process.} said \textit{P4}. 
This real-time visualization helps trainees internalize complex tasks more easily, bridging the gap between theoretical learning and hands-on practice.
Similarly, \textit{P6} noted how the structured additional information random provided could help new users: "For someone who knows nothing, the random mode gives a lot more background information, helping them understand the system." However, like the experts, \textit{P6} agreed that the \systemname would be more effective in supporting novice learners without overwhelming them, as it offers a more structured and targeted approach to learning.

Both \textit{FP1} and \textit{FP2} similarly agreed that the current structure of \systemname~is more suitable for training environments. Particularly \textit{FP2} (and feedback from \textit{I1}) noted the many individual preferences in how flight instructors teach new students. A preference-based structure to the adaptability would provide additional help to flight instructors training new pilots. \textit{FP1} noted a particular complaint in having new pilots go through paper checklists, in that, they can sometimes take excessive time to their own detriment. An adaptive feedback to help their training regime would be to notify the trainee that they are taking to long and need to move to the next step.

\vspace{-0.05in}
\myparagraph{Distinction Between Black Hawk and Leisure Aircraft}
Black Hawk helicopters operate in high-pressure, mission-critical environments where precision and quick decision-making are essential. 
These operations often involve complex checklists for preflight and emergencies. 
As \textit{E1} pointed out,\participantquote{For Black Hawk pilots, the tasks are mission-critical, and there is little room for error or distraction.}
The \systemname for Black Hawk pilots must prioritize situational awareness and provide minimalist, high-priority cues, such as engine failure warnings or critical mission navigation updates. 
\textit{E1} explained,\textit{"The adaptive system should focus only on mission-critical information, particularly during high-stress phases like combat maneuvers or landings."}
In contrast, leisure aircraft operations typically have a slower pace and lower risk, allowing pilots more flexibility to process information. \textit{E2} noted, \textit{"In leisure aircraft, the system can offer more frequent cues and continuous guidance, especially for less experienced pilots." }
During routine stages like cruising, the system can provide ongoing guidance on fuel efficiency, flight paths, or error corrections without overwhelming the pilot. 
This approach helps less experienced pilots gradually build confidence while familiarizing themselves with procedures and aircraft systems.

\vspace{-0.05in}
\myparagraph{Safety and Trust in Adaptive Flight Systems}
Both experts and participants emphasized the importance of maintaining traditional paper checklists alongside adaptive systems to ensure safety and trust during flight operations. \textit{P6} highlighted that while adaptive systems can be helpful, they should not replace the tried-and-true methods like paper checklists, especially in case of system malfunctions.
\textit{"If the plane fails, the paper checklist is still there,"} \textit{P6} noted the importance of redundancy in high-stakes situations. 

\vspace{-0.05in}
\myparagraph{Human Control is Always Necessary}
Both experts and participants agreed that adaptive systems should support, but never replace, human control.
\textit{E1} emphasized that \textit{"adaptive systems should always defer to the pilot's judgment, especially during critical phases like emergency landings or combat maneuvers."} 
This sentiment aligns with \textit{P6}'s concerns about over-reliance on automation, particularly in high-pressure situations. \textit{P6} noted that relying too heavily on the adaptive system could slow reaction times, as pilots may become accustomed to the system making decisions for them. 
In such cases, pilots may "panic" if the system fails or provides delayed feedback, making it crucial for them to remain fully engaged in controlling the aircraft.
Furthermore, experts and participants pointed out that adaptive systems should act as an assistant rather than a primary decision-maker. As \textit{P6} suggested, "The adaptive system can monitor variables and provide helpful notifications, but the pilot should always be in command." The expert interviews reinforced this, stressing that adaptive systems must assist pilots in making decisions, rather than fully controlling the aircraft. This ensures that pilots maintain situational awareness and are prepared to intervene if the system falters.
By balancing traditional tools like paper checklists with adaptive systems, and ensuring that pilots retain control over critical decisions, adaptive systems can provide effective support without undermining pilot autonomy. 
\vspace{-0.08in}
\section{Discussion}






We aimed to address the potential benefits of a neuroadaptive pilot guidance system, \systemname. The system integrated a VR simulation of a UH-60 Black Hawk modified guidance derived from the reasoning of a PHI-3 LLM model. We hypothesized that our system  would result in a higher incidence of optimal cognitive load which in turn would result in faster task completion and with less errors. However, the quantitative evaluation only partially supported this claim. We found strong evidence that relative to the checklist baseline and the random control condition an adaptive system did produce a greater incidence rate of optimal cognitive load states but only for the classifier predicting memory. This suggests the system's adaptive rules to manage memory load were successful. However, there were no appreciable differences in perception, and while we did observe a benefit regarding optimal load incidence rate for perception between the baseline and the adaptive condition, the random condition was better then both. This suggests that our strategies for managing attentional and perceptual load were ineffective. 

It is worth noting that upon examining the behavior of the LLM following cognitive load predictions regarding negative states in perception and attention the LLM was consistently confused by their differentiation and struggled to understand the appropriate guidance actions associated, specifically with attention and perception. This may have hindered the efficacy of the adaptive guidance under these classified cognitive states.
Given these results for memory, perception, and attention it was surprising to see no appreciable improvement in task completion time, and  an increase in error counts in the guidance conditions. Therefore, these results may best be interpreted in light of the additional context provided by the qualitative data we obtained from the participants after study completion.

Overall, the adaptive system was viewed favorably by participants. In the interviews the pilots emphasised \textit{the systems ability to give the right kind of information}, of the right quantity at the right time. Some emphasized the comfort the system provided in being able to guide them through the task. They even noted how the information provided by the random control while contextually relevant at times felt more detailed then was necessary. In light the experiences the pilots describe when interacting with the adaptive system, there is a real possibility that they became complacent in their execution of the preflight tasks. Here it appears that complacency manifested and reduced negative memory states, and an increase in task errors. However, it is worth noting that this increase in errors did not have a meaningful effect on task completion time. While not significant we can see that in general, the trend was for faster completion time with the adaptive system. This suggests that future adaptive systems may have to act to control task complacency which may manifest as a speed-accuracy trade-off from a shifting of criteria.

\vspace{-0.08in}
\subsection{Strategies for Future Adaptive Flight Systems}

The following is a list of strategies we have developed for future adaptive pilot guidance systems and recommendations for further studies. These strategies are based on our findings in the formative study and our case study qualitative and quantitative evaluations. It is important to note that these are not conclusions but recommendations that require further studies to prove their effectiveness.

\vspace{-0.05in}
\myparagraph{Dynamic Adaptation to Task Complexity and Cognitive State}
Neuroadaptive guidance systems should adjust feedback modalities and information density based on the pilot's task and cognitive workload. This dynamic adaptation can be beneficial for maintaining an optimal working memory state and potentially could be useful for maintaining an optimal perception state on specific tasks. Aligning with~\ref{dr:dynamic}, this approach ensures that pilots receive the right amount of information at the right time, enhancing performance without causing cognitive strain in working memory or perception. However, further analysis must be performed to conclude if this benefit in perception is from the guidance itself or the adaptation. For future studies, we recommend a further analysis of the changes in workload states based on different specific cockpit tasks.

\vspace{-0.05in}
\myparagraph{Minimize Cognitive Overload and Information Redundancy}
To prevent overwhelming pilots, especially during critical operations, guidance systems should provide concise and essential information. Excessive information can lead to cognitive overload~\ref{dr:overload}. Our findings indicated that the adaptive condition had higher error rates compared to the baseline, suggesting that too much guidance might cause a loss of focus. Participants also preferred when the system avoided unnecessary details. By minimizing redundant information, the system can help pilots maintain focus and reduce the likelihood of errors during flight tasks. Future systems should optimize different levels of information load and experiment with to what extent complacency starts and errors occur.

\vspace{-0.05in}
\myparagraph{Enhanced Error Detection and Correction Mechanisms}
Guidance systems should develop features that not only identify errors but also guide pilots in correcting them promptly and intuitively. While~\ref{dr:error}~emphasizes proactive error correction, our results showed no significant reduction in error rates with adaptive guidance. Participants indicated a need for clearer guidance when correcting mistakes, suggesting room for improvement in error recovery support. Enhancing these mechanisms will assist pilots in swiftly addressing errors, thereby improving safety and efficiency during operations. Future guidance systems should look at adaptive ways to inform users of their errors and put them back on track without disrupting optimal workload states.

\vspace{-0.05in}
\myparagraph{Enhance Trust and Reliability in Adaptive Guidance Systems}
To calibrate pilot trust and ensure robust operation, future adaptive guidance systems should maintain pilot autonomy by supporting decision-making without undermining authority or reliance on traditional methods. In alignment with design requirements, these systems must be highly reliable and automated, efficiently adjusting feedback based on cognitive states and flight phases. Incorporating redundancies prevents over-dependence and ensures functionality during failures, thereby enhancing pilot confidence and safeguarding against potential malfunctions. We recommend for future systems to be transparent with there pilots, explain when they are uncertain, and to not undermine the pilot in there recommendations.

\vspace{-0.05in}
\myparagraph{Tailoring Adaptivity to Expert and Novice Preferences}
Implement adaptive guidance systems should tailor functionalities based on the pilot's experience level, optimize guidance strategies according to specific task demands (e.g., data input vs. monitoring tasks), and allow customization of feedback modalities and levels based on individual pilot preferences and operational requirements. Our findings showed that adaptive guidance is particularly beneficial in training scenarios for novices, helping them familiarize themselves with cockpit procedures. Experienced pilots may prefer less intrusive guidance, aligning with context-specific adaptation. A future guidance system should incorporate adaptivity levels and analyze if these preferences lead to better performance.

\vspace{-0.08in}
\section{Limitations}
\revision{While the repeated-measures design and mixed-effects statistical analysis help mitigate small-\(n\) concerns, we acknowledge that having a small group of primary participants in a niche domain (pilots with UH-60 or other flight experience) limits generalizability. Accessing certified pilots for controlled experiments can be challenging due to training schedules and institutional approvals. Thus, readers should interpret our quantitative findings as indicative trends rather than definitive proofs. However, these insights still provide valuable directions for future work with larger, more diverse samples.}

\vspace{-0.08in}
\section{Conclusion}
We presented \systemname that uses a PHI-3 LLM to dynamically adapting visual, auditory, and text-based cues in response to pilots’ cognitive load during preflight operations. 
Through a formative study, we identified adaptive rules from expert pilots, informing the system's design. 
An adaptive feedback model was then developed to adjust cues in real time.
To evaluate this approach, we compared \systemname with a traditional paper checklist baseline and a random feedback system in a VR-simulated cockpit environment. 
Results show that \systemname accelerated task completion time relative to the baseline and random system condition. This was accomplished with in conjunction with greater rates of optimal memory and perception relative to the baseline. Overall, we demonstrated that  \systemname has the potential to enhance pilot performance, especially in complex cockpit environments. However, further research is needed to refine, test, develop and tune of adaptive systems. Specifically in terms of how they perceive cognitive load, how they reason over the application of adaptive strategies and the nature of the adaptive strategies they use.

\section{SUPPLEMENTAL MATERIALS}
All supplemental materials are available at \url{https://github.com/ShaoyueWen/AdaptiveCopilot} released under a CC BY 4.0 license. In particular, they include (1) Figures for system design, (2) Figures for false error rates, (3) Preflight paper checklist, (4) Formative study questions (5) Qualitative evaluation questions, (5) Prompt Examples, (6) Adaptive rules. \revision{Certain proprietary assets (trained cognitive models, Unity simulation) prevent us from open-sourcing the system in its entirety. Requests for access to the dataset or code base should be directed to the corresponding author.
}

\vspace{-0.08in}
\section {Acknowledgments}
Approved for Public Release; NG24-2234; @2024 Northrop Grumman Systems Corporation.  This work was supported by the DARPA PTG program.  Any opinions, findings, and conclusions or recommendations expressed in this material are those of the authors and do not necessarily reflect the views of DARPA.

\bibliographystyle{abbrv-doi}

\bibliography{references}

\appendix

\end{document}